\documentstyle[times,epsfig]{mn}
\input{epsf}  
 \newbox\grsign \setbox\grsign=\hbox{$>$} \newdimen\grdimen \grdimen=\ht\grsign  
\newbox\simlessbox \newbox\simgreatbox \newbox\simpropbox  
\setbox\simgreatbox=\hbox{\raise.5ex\hbox{$>$}\llap  
     {\lower.5ex\hbox{$\sim$}}}\ht1=\grdimen\dp1=0pt  
\setbox\simlessbox=\hbox{\raise.5ex\hbox{$<$}\llap  
     {\lower.5ex\hbox{$\sim$}}}\ht2=\grdimen\dp2=0pt  
\setbox\simpropbox=\hbox{\raise.5ex\hbox{$\propto$}\llap  
     {\lower.5ex\hbox{$\sim$}}}\ht2=\grdimen\dp2=0pt  

\def\simlt{\mathrel{\copy\simlessbox}}  

\begin{document}

\title[Reflection at large distance in Seyferts]{Reflection at large distance from the central engine in Seyferts}

\author[J.~Malzac, P.O.~Petrucci]{\parbox[]{6.8in}{ 
{Julien~Malzac$^{1}$, 
Pierre-Olivier~Petrucci$^{1,2}$}}\\
$^{1}${Osservatorio Astronomico di Brera, via Brera, 28, 20121 Milan, Italy}\\ 
$^{2}${Laboratoire d'Astrophysique, Observatoire de Grenoble, BP 53X, 38041 Grenoble Cedex, France}}

\date{Accepted, Received}  
 
\maketitle

\begin{abstract}
We consider the possibility that most of the reflection component,
observed in the hard X-ray spectra of Seyfert galaxies, could be formed
on an extended medium, at large distance from the central source of
primary radiation (e.g. on a torus). Then, the reflector cannot respond
to the rapid fluctuations of the primary source. The observed reflected
flux is controlled by the time-averaged primary spectrum rather than the
instantaneous (observed) one. We show that this effect strongly
influences the spectral fit parameters derived under the assumption of a
reflection component consistent with the primary radiation.  We find that
a pivoting primary power-law spectrum with a nearly constant Comptonised
luminosity may account for the reported correlation between the
reflection amplitude $R$ and the spectral index $\Gamma$, and
simultaneously produces an iron line $EW$ that is nearly independent of
$\Gamma$.  We emphasize the effects of the modelling of the primary
component on the determination of the reflection amplitude, and show that
in NGC 5548, when these effects are taken into account, the {\it RXTE} data are
consistent with the reflection features being produced mainly at large
distance from the central source.
\end{abstract}

  
%################################################################  

\footnotetext{E-mail: malzac@brera.mi.astro.it} 
 
\section{Introduction}
 
In radio-quiet active galactic nuclei, large amounts of cold material are
thought to reside at large distance from the central engine. In
particular, the unified scheme (Antonucci 1993) postulates a large scale
torus (with radius larger than $10^{16}$ cm).  Such {distant cold
material (hereafter DCM)} may imprint strong reflection features in the
hard X-ray spectrum of Seyfert galaxies.  These spectral characteristics
take the form of a Compton hump in the X-ray spectrum (e.g. Lightman \&
White 1988; George \& Fabian 1991) and a narrow iron line component at
6.4 keV. They add to similar features likely to form much closer to the
central source presumably by reflection of the primary X-ray radiation on
the inner parts of the accretion disc.

In Compton thick Seyfert 2s, the observed reflection dominated spectra
are generally interpreted by DCM reflection with the primary emission
obscured presumably by the same outer material (e.g. Reynolds et al.
1994; Matt et al. 2000). In Compton thin Seyfert 2s, Risaliti (2002)
recently present, from the analysis of a large number of {\it BeppoSAX}
observations, strong evidences supporting the presence of a distant
reflector in these objects.  The unifications models postulate that such
material is also present in Seyfert~1s but out of the line of sight.  In
this context, it has been argued that the reflection produced by an
irradiated torus would be sufficient to account for the typical observed
reflection spectra in Seyfert 1s (Ghisellini, Haardt \& Matt 1994;
Krolik, Madau \& \.Zycki 1994), without need for reflecting material in
the inner parts of the accretion flow.

The first evidence for remote reflection in a type 1 Seyfert came from
the {\it BeppoSAX} observation of NGC 4051 in a ultra faint state where
the nuclear emission had switched-off leaving only a spectrum of pure
reflection from material at distance larger than 10$^{17}$ cm (Guainazzi
et al. 1998; Uttley et al. 1999).  However, the presence of a reflecting
torus in all Seyfert 1s is still a matter of debate.

This important issue motivated the search for a narrow iron emission line
that would be the signature for reflection on distant cold and
Compton-thick material.  Recent {\it Chandra} (e.g. Kaspi et al. 2001; Yaqoob
et al. 2001) and {\it XMM-Newton} (e.g. Reeves et al. 2001; Pounds et al. 2001;
Gondoin et al. 2001a, 2001b; Matt et al. 2001, O'Brien et al. 2001)
observations indicate that a narrow line is often, if not always, present
in Seyfert 1 galaxies, alone or together with a relativistic line
component (Tanaka et al. 1995; Fabian et al 2000 and reference therein).
An analysis of the composite ASCA spectrum of Seyfert 1s (Lubi\~nski \&
Zdziarski 2001, hereafter LZ01) also suggests the presence of a narrow
 component confirming previous findings on individual sources.

An important complementary approach in order to locate the place of
emission of the different components is the study of the variability of
the reflected features.  Indeed, the temporal response of the line to
fluctuations of the continuum depends on the size and distance of the
reflector.  Reflection features formed in the disc, in the broad line region,
 or in a torus are expected to show little variability on time scales shorter
 than a few hours, a few weeks or a few months respectively.  In this framework
the monitoring campaigns with {\it RXTE} of several objects gave
indications for a constant line and reflection component flux while the
continuum was strongly variable (Papadakis et al. 2002).

There is thus growing evidence in favor of a significant fraction of the
reflection features being formed at large distance from the central
engine.

On the other hand there is an important  
observational and theoretical issue 
that is the existence of a correlation between the amplitude of the
 reflection component $R$ and the spectral index $\Gamma$ 
found by Zdziarski, Lubi\~{n}ski \& Smith 1999 (hereafter ZLS99) 
in a large sample of {\it Ginga} spectra.
 The measured $R$
tends to be larger in softer sources. This correlation is observed 
in sample of sources as
well as in the time evolution of individual sources. 
 The $R$-$\Gamma$ correlation is
also found in the {\it BeppoSAX} data (Matt 2001), the {\it BeppoSAX}
sample showing however a significantly different shape, with on average
higher $R$ values at low $\Gamma$ and a flatter correlation.
The usual interpretation of the correlation invokes the feedback from
reprocessed radiation emitted by the reflector itself. It thus requires
 the main source of reflection to be present in the direct environement 
of the $X$-ray source. This thus appears inconsistent with the DCM picture. 

One of the main goal of this paper is to attempt solving this paradox.
We will show that under assumptions 
based on the present knowledge 
of the Seyfert 1s phenomenology, a $R$-$\Gamma$
 correlation from a distant reflector is actually expected. 
Another possibility is that the correlation could be spurious.
The $R$-$\Gamma$
 correlation has been indeed widely controversed, in particular regarding 
the uncertainties on the measurement of $R$, but no-one could prove 
unambiguously that it is an artifact.
 In this paper we demonstrate that the determination of $R$ is sensitive
 to  the assumed high energy cut-off and this may strongly 
alter the shape of the correlation.

In section~\ref{sec:qualeffects}, we recall
the generic effects of the DCM reflection on the measured amplitude of
the reflection features. In section~\ref{sec:qeffects}, we investigate
quantitatively the dependence of the reflection amplitude and the iron
line equivalent width (EW) on the spectral index $\Gamma$ in the case
where the variability of the primary component can be described as a
pivoting power-law.  Finally, in section~\ref{sec:ngc5548} we attempt to
test the presence of DCM reflection in NGC 5548 using {\it RXTE} data.

\section{Qualitative effects of a distant reflector}\label{sec:qualeffects}

As the DCM line is spectrally distinct from the broader line emitted
close to the black hole, the existence and the distance/scale of the DCM
line  region can, in principle, be probed using the new high
resolution X-ray telescopes, as discussed above.  On the other hand, the
DCM Compton hump is more difficult to measure accurately and disentangle
from its similar disc counterpart.

Actually, when analyzing the X-ray data, the possible contribution from
DCM reflection is usually not considered a priori.  The data are
generally interpreted in the framework of reflection in the vicinity of
the hard X-ray source.  Usually, in
spectral fits, the shape of the reflection component is computed assuming
that the \emph{observed} primary spectrum illuminates an infinite slab
(e.g. {\sc pexrav} model in {\sc xspec}, Magdziarz \& Zdziarski 1995).
The normalization of the reflected spectrum is then tuned in order to fit
the observed spectrum.  The result of this fitting procedure provides the
reflection amplitude $R$. $R$ is normalized so that $R=1$ in the case of
an isotropic source above an infinite reflecting plane. $R$ is often
considered as an estimate of $\Omega/2\pi$ where $\Omega$ is the solid
angle subtended by the reflector as seen from the isotropic X-ray source.
Obviously, any contribution from a remote structure leads to an increase
of $R$ and may lead to an overestimate of the disc reflection.  This may
explain the very large $R$ coefficients $\sim 2$, measured in some
Seyfert 1s, which are difficult to reconcile with disc reflection (see
however Beloborodov 1999; Malzac, Beloborodov \& Poutanen 2001).

The nuclei of Seyfert galaxies are known to harbor a significant flux and
 spectral variability on very short time-scales (see e.g. the recent
 review by Nandra 2001).
Due to its extended structure, the remote reflector cannot respond to the
rapid fluctuations of the primary X-ray flux.  \emph{The reflected
component from the DCM is thus likely to correspond to the time-averaged
incident flux seen by the cold material rather than to the instantaneous
(i.e. observed) one.}  Thus flux changes may induce a significant
variation in the $R$ value derived from the spectral fits.  A flux lower
than the average enhances the apparent reflection, on the other hand, a
larger flux may reduce $R$ down to zero.  Then values of $R$ as low or
large as required by the data can be easily produced. Similarly, the DCM
iron line flux is expected to be constant and, consequently, its
equivalent width would vary like the inverse of the primary continuum
flux around the line energy.

\begin{figure} % fig.1
\vspace{10pt}
\centerline{\includegraphics{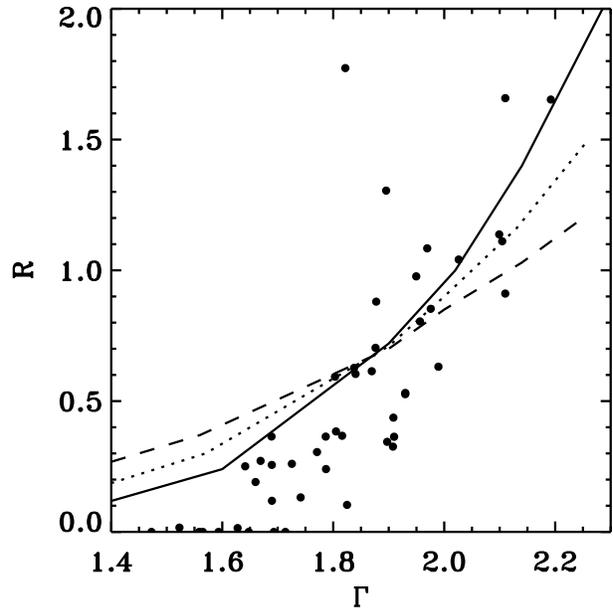}}
\caption{The curves show the $R$--$\Gamma$ correlation obtained assuming
a primary power-law spectrum pivoting around 2 (solid), 5 (dots) and 10
keV (dashes) with a fixed exponential cut-off at 400 keV and a fixed
reflected flux and spectrum corresponding to the reflection $R=0.7$ and
$\Gamma$=1.9.  The simulated spectra were fitted using {\sc pexrav} in
the 2-30 keV range.  The circles show the {\it Ginga} data of Zdziarski et
al.(1999).  The errors have been omitted for clarity.}
\label{fig:smallfig1}
\end{figure}

\begin{figure*} % fig.2
\vspace{10pt}
\centerline{\includegraphics{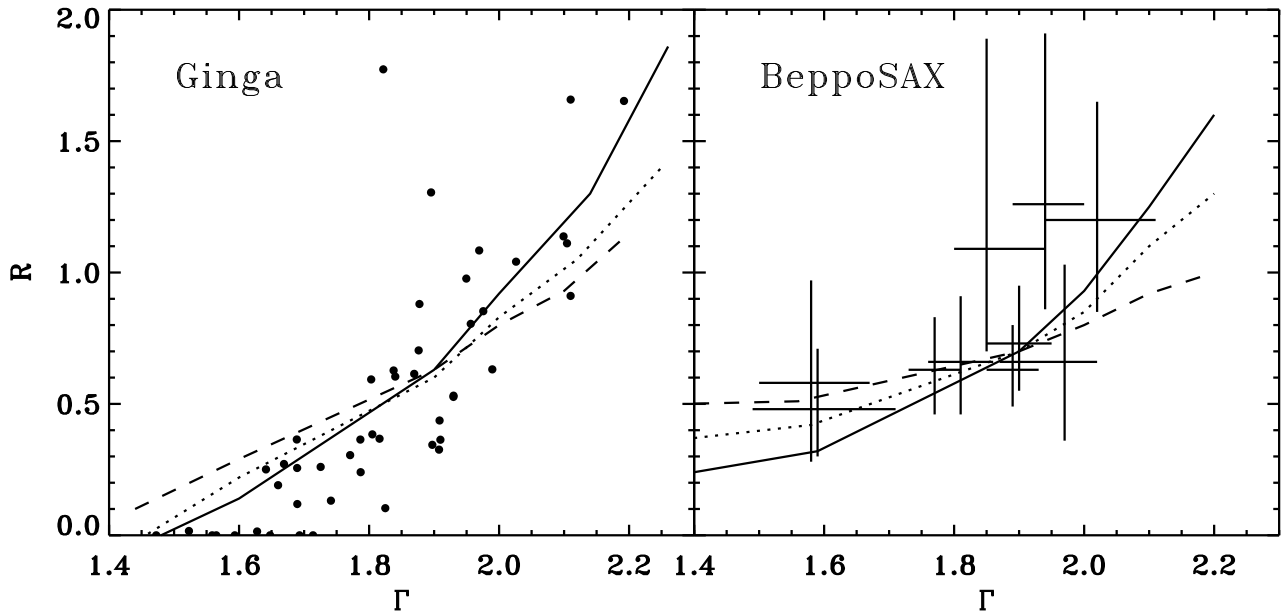}}
\caption{The curves show the $R$--$\Gamma$ correlation obtained assuming
a primary powerlaw spectrum pivoting around 2 (solid), 5 (dots) and 10
keV (dashes) with an exponential cut-off depending on $\Gamma$ according
to the relation of equation~\ref{eq:ecgamcorr}, and a fixed reflected flux and
spectrum corresponding to the reflection $R=0.7$ and $\Gamma$=1.9. Left
panel: the simulated spectra were fitted using {\sc pexrav} in the 2-30
keV range.  The circles show the {\it Ginga} data of Zdziarski et al. (1999)
and the errors have been omitted for clarity.  Right Panel: the simulated
spectra were fitted using {\sc pexrav} in the 2-200 keV range.  The
crosses show the {\it BeppoSAX} data of Perola et al. (2002)}
\label{fig:smallfig2}

\end{figure*}

As quoted in the introduction,
 a significant contribution from a remote reflector seems,
at first sight, in contradiction with the reported correlation between
$R$ and the spectral slope $\Gamma$. 
Indeed, in the context of DCM reflection, it is difficult to
understand why the reflection contribution from DCM should be more
important in objects with softer spectra.
However, beside the simple effects of flux variability at constant spectrum
 discussed above, the presence of \emph{spectral} variability also
 alters the appearance of the DCM reflection.
 Spectral variations with $\Delta\Gamma$$>$0.1
are indeed commonly reported in Seyfert 1s.
 In the case of reflection on DCM,
such spectral changes 
 have a significant impact on the measured amplitude 
of the reflection features and the measurement of $R$.
The resulting $R$ vs $\Gamma$ relation
then depends on the specific spectral variability mode of the sources that may,
 under certain circumstances, produce a $R$-$\Gamma$ correlation
 (Nandra et al. 2000).

It appears indeed that the observed variability in Seyfert galaxies
 is generally consistent with fluctuations of the X-ray spectral slope
 with a nearly constant comptonised luminosity, i.e.
 the spectrum is mainly pivoting.
This can happen for example if the UV luminosity entering the hot
Comptonising plasma changes with a constant heating rate in the hot
plasma (e.g. Malzac \& Jourdain 2000).  This kind of behaviour is widely
 observed in numerous sources such as NGC 5548 (Nicastro et al. 2000;
Petrucci et al. 2000), 3C 120 (Zdziarski \& Grandi 2001), NGC 7469
(Nandra et al. 2000; Nandra \& Papadakis 2001), 1H 0419-577 (Page et
al. 2002), NGC 3783 (De Rosa et al. 2002). 
In general, the pivot point appears to be located in the 2-10
keV range (see fig.~9 of Petrucci et al. 2000, fig.~3 of Page et
al. 2002, fig.~3 of Zdziarski \& Grandi 2001).

In the case of a pivoting spectrum at energies lower than 10 keV -- i.e.
below the energy range were most of reflection is produced -- when the
spectrum is hard the observed primary flux in the 10--30 keV band is
larger, and the relative amplitude of the DCM reflection, $R$, is reduced.
  On the other hand when the
spectrum is soft the primary flux in this energy range is enhanced and
the measured $R$ is larger.  This thus produces \emph{a positive
correlation between $R$ and $\Gamma$}.  The lower is the energy of the
pivoting point, the steeper the correlation will be.

\section{Quantitative effects of a distant reflector}\label{sec:qeffects}

We now attempt to investigate more quantitatively the effects of
 a distant reflector on the relation between the spectral index $\Gamma$ and
 the iron line equivalent width and reflection amplitude $R$.

As noted in the previous section, the $R$-$\Gamma$
 relation depends on the variability 
mode. If one had a very accurate measurement of the $R$-$\Gamma$ correlation
we could derive the variability mode required in order to produce 
the observed correlation with a DCM. However due to the large uncertainties
 in the measurement of $R$ and $\Gamma$ this is of little interest.
 We thus adopted the opposite approach:
 given the main observed characteristics 
of the variability in Seyfert 1s  
we test wether the presence of DCM can produce a correlation that 
is qualitatively similar to that observed.
As discussed above, it seems that
 the spectral variability, 
in general, can be, at least roughly, described in term of a pivoting 
spectrum. We will thus assume that this pivoting spectrum represent
 the general (or average) variability mode of the Seyfert galaxies.
We caution however that the details of the variability mode in Seyferts is  
not known, and the single pivoting point is only a simplifying approximation.

\subsection{Model with constant cut-off energy}\label{sec:cstec}

We assume that the time average primary spectrum seen by the DCM can be
represented by a power law with a photon index $\Gamma=1.9$ --
corresponding to the average $\Gamma$ in Seyferts 1 galaxies (Nandra et
al. 1994; Matt 2001) -- and an exponential cut-off at 400 keV. Neglecting
any disc reflection, we further assume that the DCM geometry is such that
the time averaged primary spectrum yield a reflection coefficient $R=0.7$
(close to the average observed $R$ for sources with $\Gamma\sim1.9$) .  As a
first order approximation, the shape of the reflection spectrum is
computed using the {\sc pexrav} procedure, i.e. assuming a slab
reflector.  Fixing the cut-off energy at 400 keV, we produced a set of
instantaneous primary spectra for several photon indices $\Gamma$
spanning the observed range 1.4--2.2.  We did this for pivot points at
energies 2, 5 and 10 keV respectively.

We then added the reflection component, produced as described above, to
these instantaneous spectra, and fitted the resulting spectra with {\sc
pexrav} in the 2-30 keV range.  We also fitted the simulated spectra in
the 2--100 keV range. In both energy ranges, the derived best fit
parameters were very similar.  The $R$-$\Gamma$ relations resulting from
our modelling procedure are shown in Fig.~\ref{fig:smallfig1} and compared
with the correlation observed in the {\it Ginga} \, data.

Although the predicted correlation appears flatter than the observed one
it \emph{qualitatively matches the $R$-$\Gamma$ correlation observed in
the sample of {\it Ginga} data from ZLS99}.

\subsection{Model with variable cut-off energy}\label{sec:varcut}

Actually the amplitude of the reflection component is difficult to
measure accurately and the derived $R$ values are quite dependent on the
modelling of the primary emission (see Petrucci et al. 2001).  In
particular, the {\it Ginga} data used by ZLS99 to demonstrate the correlation
do not enable one to constrain the cut-off energy.  In their analysis
they thus fixed $E_{c}$=400 keV in all their fits.  Actually the $R$
derived from spectral fits depends on this assumed $E_{c}$ value. The
{\it BeppoSAX} data allowed better constraints on the cut-off energy.
 Fitting these data with {\sc pexrav } 
shows that the cut-off energy differs from source to source and is also
apparently correlated with the spectral slope.
 Fitting the $E_{c}$-$\Gamma$ correlation presented by Perola et al.  (2002),
we found that this correlation can be qualitatively represented by the
following function:
\begin{equation}
E_{c}= 10^{0.73}\,\Gamma^{5.7}\qquad\rm{keV}.
\label{eq:ecgamcorr}
\end{equation}
Perola et al. (2002) also suggested that the effects of the different
cut-off energies could be responsible for part of the already mentioned
 differences between the {\it BeppoSAX} and {\it Ginga} \, 
$R$--$\Gamma$ correlations.

In order, to check the effect of a varying cut-off on the correlation
produced by a DCM reflection, we generated a set of simulated spectra in
a way similar to that described in section~\ref{sec:cstec}, but instead
of a constant $E_{c}$, we introduced a ``$\Gamma$--dependent'' $E_{c}$ as
given by equation~\ref{eq:ecgamcorr}.
 
First, we mimicked the {\it Ginga} \, analysis of ZLS99 by fitting the set of
simulated spectra in the 2-30 keV range assuming a constant cut-off at
$E_{c}$=400 keV.  As shown in the left panel of Fig.~\ref{fig:smallfig2},
this produces $R$-$\Gamma$ correlations which are steeper than in the
case of the models with constant $E_{c}$.  In particular at low $\Gamma$,
the reflection amplitude is lower and reaches $R$=0 for the hardest
spectra, as indeed observed by ZLS99.  Thus allowing a correlation
between $E_{c}$ and $\Gamma$ results on a better agreement of the DCM
model with the observations.

We then fitted the simulated spectra in the broader (2--200 keV) energy
range that is allowed by {\it BeppoSAX}. The cut-off energy $E_{c}$ was then
let as a free parameter and we consistently recovered best fit $E_{c}$
close to the injected ones.  This procedure results in $R$-$\Gamma$
correlations which are flatter and with higher $R$ values, consistent
with the {\it BeppoSAX} correlation as can be seen from the right panel of
Fig.~\ref{fig:smallfig2}.

We note that the $E_{c}$-$\Gamma$ correlation found with {\sc pexrav}
 in the {\it BeppoSAX}
 data does not necessarily reflect changes in the temperature of 
the Comptonising plasma. Indeed, although 
 {\sc pexrav} is useful to provide a simple and standard phenomenological 
description of observed spectra, this model represents only a rough
 approximation to real Comptonisation spectra. The 
physical interpretation of the fit parameters $\Gamma$ and $E_{c}$
in terms of Thomson optical depth and temperature is not straightforward.  
Moreover the use of {\sc pexrav} in spectral fits may produce
 or emphasize parameter correlations that are not necessarily found 
with realistic Comptonisation models (Petrucci et al 2000, 2001).
To see if the $\Gamma$-$E_{c}$ correlation obtained with {\sc pexrav}
 could be produced at constant temperature, we used the {\sc compPS} model\footnote{available at ftp://ftp.astro.su.se/pub/juri/XSPEC/COMPPS/} 
(Poutanen \& Svensson 1996) to generate a set of realistic Compton spectra
that we subsequently fit with  {\sc pexrav}.
In the {\sc compPS} simulations we assumed a spherical and isotropic
 geometry for the comptonising plasma with temperature fixed at
$kT_{e}$=100 keV, and 
varied the Thomson optical depth of the comptonising 
plasma. We added to all spectra the same reflected flux. 
We fit these spectra with {\sc pexrav} in the 2--200 keV range 
and indeed obtained a correlation between the resulting
  $E_{c}$ and $\Gamma$ parameters.
For $\Gamma< 1.9$, this correlation is 
however flatter than the  observed $E_{c}$-$\Gamma$ relation given by 
equation~\ref{eq:ecgamcorr}.
The physical significance (if any) of the  $E_{c}$-$\Gamma$ 
correlation is thus unclear. From the phenomenological side however, this 
correlation provides the best {\sc pexrav} representation of the data.
It thus should be taken into acount when simulating observed spectra
 with this model.

We further note that the 
$R$-$\Gamma$ correlation resulting from our  {\sc compPS} simulations
 with  constant temperature, when these spectra are fit with {\sc pexrav}
 in the {\it Ginga} range, is intermediate between what obtained
 for a constant $E_{c}$ (Fig.~\ref{fig:smallfig1}) and for
$E_{c}$ given by equation~\ref{eq:ecgamcorr}
 (left panel of Fig.~\ref{fig:smallfig2}).

\begin{figure} % fig.3
\vspace{10pt}
\centerline{\includegraphics{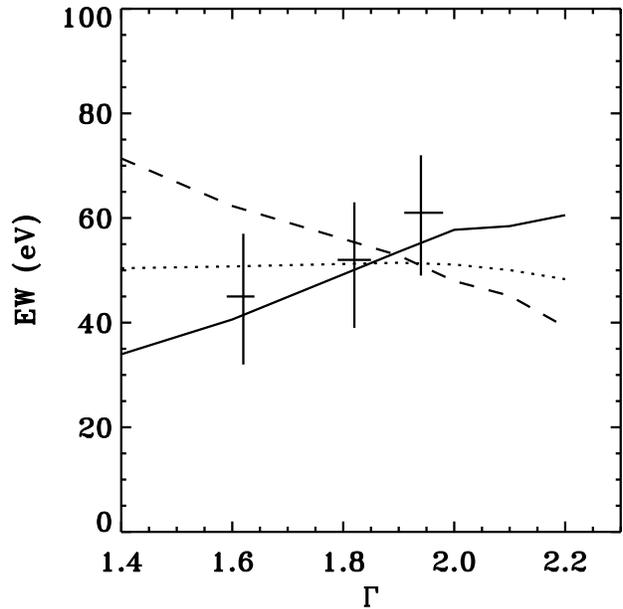}}
\caption{Curves: the $EW$--$\Gamma$ relation obtained assuming a primary
power-law spectrum pivoting around 2 (solid), 5 (dots) and 10 keV (dashes)
with a fixed exponential cut-off at 400 keV and a fixed line flux (see
text).  The simulated spectra were fitted using {\sc pexrav} in the 2-30
keV range.  The crosses show the narrow line EW from the average ASCA
spectra of Lubi\~{n}ski \& Zdziarski (2001).}
\label{fig:smallfig11}
\end{figure}    

\begin{figure*} fig.4

\centerline{\includegraphics{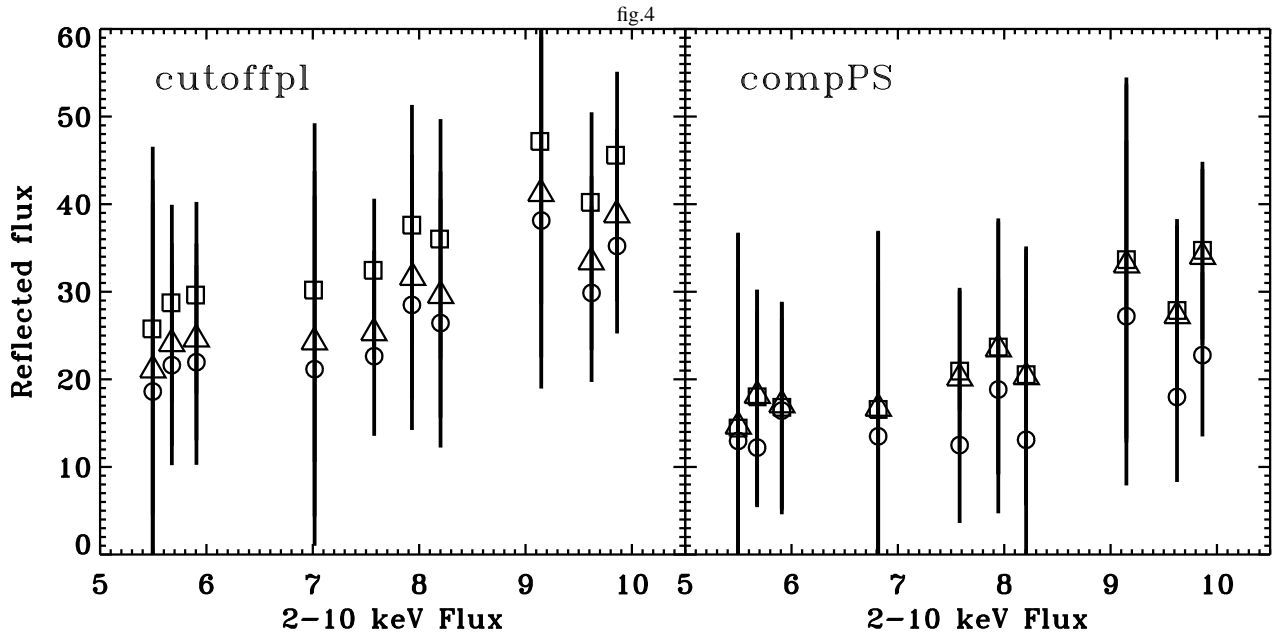}}
\caption{The reflection flux versus 2-10 keV flux in NGC 5548.  Left
panel: the primary emission is modeled using an e-folded power-law ({\sc
cuttoffpl} model). Squares, triangles and circles represent the best fit
values assuming a constant cut-off energy $E_{c}$ at 100, 200 and 400
keV respectively.  Right panel: the primary emission is modeled using a
Comptonisation spectrum ({\sc compPS} model) assuming a spherical
geometry.  Squares, triangles and circles stand for fixed comptonising
plasma temperature $kT_{e}$=40, 160 and 300 keV respectively.  The 2-10
keV and reflected flux are both given in units of 10$^{-11}$ ergs s$^{-1}$ 
cm$^{-2}$. The error bars are at the 90 per cent confidence level}
\label{fig:rfluxchiang}
\end{figure*}

\subsection{The iron line equivalent width vs $\Gamma$ relation}

We now discuss the effects of a pivoting spectrum on the measured Fe line
equivalent width ($EW$) produced by the DCM.  For a power-law spectrum
pivoting at an energy $E_{p}$, and a pure DCM spectrum, we have:
\begin{equation}
EW=\frac{I_{Fe}}{F_{E_{p}}}\left(\frac{E_{p}}{E_{K \alpha}} \right)^{-\Gamma},
\end{equation}
with $I_{Fe}$ being the line photon intensity and $F_{E_{p}}$ the primary
photon flux at energy $E_{p}$.  From this relation it appears the $EW$ is
positively correlated with $\Gamma$ when the pivoting point is below
$E_{K\alpha}\simeq 6.4$ keV, while it is anti-correlated for higher
pivoting energy (the $EW$ indeed represents the amount of reflection at
6.4~keV i.e. at a lower energy than the reflection bump).  Of course, the
normalization depends on the geometry of the DCM reflector. For an
optically thick torus with a 30 $\deg$ opening angle, Ghisellini, Haardt
and Matt (1994) derive an equivalent width $EW\sim 80$ eV.  This estimate
depends also widely on viewing angle and elemental abundances (\.Zycki \&
Cerzny 1994; George \& Fabian 1991).  Using a large ASCA sample,
LZ01 produced 3 average spectra of Seyfert 1s
grouped according to increasing spectral index. For each spectra they
could detect a narrow Fe line component that they attribute to reflection
on a torus.  Their estimates for the narrow line EW are shown in
Fig.~\ref{fig:smallfig11}. It appears to be almost independent of
$\Gamma$ with an average EW of 50 eV.

Using a procedure similar to that described in section~\ref{sec:cstec}
and \ref{sec:varcut}, we simulated the effects of spectral variability on
the measured equivalent width (EW) of the iron line.  We modeled the
 DCM iron line by a Gaussian with an intrinsic width
$\sigma=0.1$ keV. The line flux was normalized to the observational
result of LZ01, i.e. so that the $EW$ is 50 eV for the $\Gamma=1.9$ and
$R=0.7$ time-averaged primary spectra.

The resulting dependence of the measured equivalent width with $\Gamma$
is plotted on Fig~\ref{fig:smallfig11}.  We see that the trend
(correlation or anti-correlation) in the $EW$-$\Gamma$ relation is indeed
qualitatively sensitive to the pivot point energy.  However, if one
consider that the pivot point location is likely to fluctuate (with time
or from source to source), our assumed variability mode predicts
essentially no specific relation between $EW$ and $\Gamma$.  Moreover,
the fluctuations of $EW$ due to a pivoting spectrum are quite weak. Thus
for a fixed geometry, our pivoting spectrum will produce an essentially
constant $EW$, independently of $\Gamma$, as observed by LZ01.

We note that the results of the averaging procedure of ZL01 should be
taken with a caution (see Yaqoob et al. 2002).  On the other hand,
accurate measurements of the $EW$ of the narrow line in individual
sources is somewhat difficult since it requires a high resolution
observations and a good determination of the continuum and the broad-line
component.  Actually, the present estimates have large error bars and
seem to depend on the modelling of the broad-line component and in
particular, on the choice of the unconstrained parameters of the
relativistic disc model.  For instance, using {\it XMM-Newton} data of Mrk 359,
O'Brien et al. (2001) find an $EW$ of the narrow line of 120 eV when
assuming a disc extending down to 3~Rs, and $EW=62$~eV when fixing the
disc inner radius at 50~Rs. Despite these uncertainties, the data on
individual sources suggest that the $EW$ of the narrow line changes from
source to source and/or on long time scale (i.e. longer than a few
years).  Although an anti-correlation between the narrow line $EW$ and
source luminosity is reported (Nandra et al. 1997; Pound \& Reeves
2002), the changes in the narrow line strength are apparently unrelated
to $\Gamma$. 

Actually, the observed fluctuations of the narrow line $EW$ are likely to
be dominated by intrinsic differences for instance in the geometry,
elemental abundances or viewing angle of the reflecting material.  Such
differences are smeared out by the averaging procedure of LZ01 that shows
the average $EW$-$\Gamma$ (in)dependence.

Thus, our assumed pivoting spectrum induces only weak changes in the
narrow line $EW$ that are observationally dominated by both uncertainties
on the line strength measurement and a spread in the characteristics of
the reflector from source to source.  Simultaneously, the pivoting
spectrum produces a strong dependence of $R$ on $\Gamma$ that would
dominate over the intrinsic differences from source to source 
and may contribute significantly
to the reported $R$-$\Gamma$ correlation.  In this context, part of the
observed spread in the $R$-$\Gamma$ plane could be due to the slightly
different DCM configurations in different sources. 
 The prediction of the absence of strong correlation between
the narrow line EW and $\Gamma$
 (linked with our assumed variability mode) is thus consistent with
 the present data and will be possibly tested when better data
 will be available.

\section{The case of NGC 5548}\label{sec:ngc5548}

From the previous section we see that the effects of DCM reflection,
when important, make the geometrical interpretation of $R$ extremely
misleading.  \emph{Trying to disentangle the temporal and geometrical
effects is extremely difficult} and requires very long observations with
time resolved spectral analysis.  NGC 5548 was the target of such a
monitoring campaign in 1999.  It was observed in the X-rays on June
15-16, June 19-24, June 29-July 2, Jul 7 and August 16-18 with {\it RXTE}, thus
providing a relatively good coverage of the source evolution during
nearly 2 months.  The results of these observations (as well as other
contemporary multi-wavelength observations) are reported in Chiang et
al. (2000, hereafter C00) .  Interestingly, while the 2-10 keV flux
changed by a factor of 2 during the campaign, C00 report a constant line
flux suggestive of a DCM line.  Strikingly the reflection amplitude $R$
was found to be essentially constant, with a trend for a correlation
between $R$ and the 2-10 keV flux.  The latter result suggests a close
origin for the reflection bump and this is clearly inconsistent with DCM
reflection.  Then the line and the reflection bump should have a
different origin, which appears difficult to explain.

In the case where the reflection component arises mainly from a distant
reflector, the $R$ parameter is not very convenient to provide a physical
insight.  Rather the absolute reflection flux may be more relevant to
probe the distant reflector since it is expected to be constant on short
time scales (a few months) independently of the source flux and spectrum.
On the other hand, in the case of close-by reflection, the reflected flux
is expected to follow the fluctuations of the primary flux.  We thus
reanalyzed the C00 data, with a model of constant reflection spectrum
using as fitting parameter, the absolute reflection flux $RF$ instead of
the usual reflection amplitude $R$.

We used the 10 {\it RXTE} spectra taken at different epochs described in C00.
Instead of using directly {\sc pexrav} as did C00, our model consisted in
an e-folded power-law ({\sc cutoffpl}) plus a reflection component with a
fixed shape (i.e. independent of the primary spectrum parameters). This
reflection spectrum was produced by {\sc pexrav} for a spectral index
$\Gamma=1.9$, a cut-off at 200 keV and standard abundances.  The absolute
normalization of the reflection component was a free parameter.  As the
reflection spectral shape is identical in all fits, this parameter is
directly proportional to $RF$. The {\it RXTE} data do not enable one to
constrain the high energy cut-off and we had to keep it frozen at an
arbitrary value. We performed three fits of each spectrum assuming
$E_{c}$=100, 200 and 400 keV respectively.
 
The dependence of $RF$ on the 2-10 keV flux is displayed on
Fig.~\ref{fig:rfluxchiang}.  For a fixed value of the cut-off energy the
data are formally consistent with a constant value of $RF$. When the RF
vs 2-10 keV flux relation is fit with a constant, the reduced $\chi ^{2}$
values are all below unity and the $\chi ^{2}$ probability is
systematically larger than $0.5$.  However, when a linear relation
between the reflection flux and the 2-10 keV flux is assumed, we
systematically get an improvement of the fit significant at better than
the 95 per cent confidence level according to $F$ test.  This correlation
between the reflection normalization and 2--10 keV flux (in agreement with the finding of
C00) is suggestive of a close reflector where the reflection flux
increases almost linearly with the primary X-ray flux.

Nevertheless, as can be seen on Fig.~\ref{fig:rfluxchiang}, the measured
$RF$ appears to depend on the cut-off energy.  For our three fixed
$E_{c}$ values we find 3 best fit $RF$ differing substantially.  The $RF$
is systematically larger for lower $E_{c}$, the typical differences being
about 40 per cent between $E_{c}=400$ keV and $E_{c}=100$ keV.  Now, if we
consider the uncertainties on the cut-off energy it appears that the
observed $RF$ is also consistent within the error bars with a DCM
reflection (i.e. constant $RF$).  In particular one may consider that the
actual $E_{c}$ might be positively correlated with the 2--10 keV
flux.

Actually, the effects of the assumed $E_{c}$ on the $RF$ could be an
artifact of our model associated with the particular e-folding cut-off
shape assumed which may differ from a real thermal Comptonisation
cut-off. We thus also model the primary emission using the more
physical Comptonisation model {\sc compPS}.
We used the Thomson optical depth as the fitting parameter. Each
spectrum was fit successively with the hot cloud temperature frozen at
three different test values (40, 160, and 300 keV).  On average, we
obtain significantly lower $RF$s with the comptonisation model.  The
determination of the reflection is indeed very sensitive to the modelling
of the primary continuum (see Petrucci et al. 2001). More precisely, 
as long as the
temperature was fixed to a low value ($\simlt$160 keV) the best fit $RF$
appears to be independent on the assumed temperature.  There is a
correlation between $RF$ and 2-10 keV flux (significant at the 98~per cent
confidence level) consistent with a close reflector, although the DCM can
not be formally ruled out.  On the other hand at higher temperature it
appears that $RF$ becomes strongly dependent on the assumed $T_{e}$. For
$T_{e}$$=$300 keV we found significantly lower values for the $RF$ and
almost no-correlation with the 2-10 keV flux. The improvement of the fit
between a constant and linear model is significant only at the 63~per cent
confidence level.  This difference with respect to the low temperature
case, is due to the fact that at large temperature the best fit optical
depth is in general quite low ($\tau_{T}\sim0.2$).  The Comptonisation
spectrum then differs from a pure power-law and this affects the measured
$RF$.

The main conclusion here is that
 whatever the fitting model used, the measured reflection amplitude is
very dependent on the assumptions on the cut-off energy (or temperature)
which can be constrained, together with the reflection component, only
with broad band instrument such as {\it BeppoSAX} or the future {\it INTEGRAL}.

As shown above, when these uncertainties are considered 
the present data become consistent with the DCM model.
In particular, our results are fully consistent with a DCM reflector in
the case of a high temperature Comptonising plasma, or a cut-off energy
positively correlated with the 2-10 keV flux. It is worth noting that 
such high
temperatures of order of 300 keV are indeed deduced from spectral fits to
the the {\it BeppoSAX} data with realistic Comptonisation models 
(Petrucci et al. 2000) and that a correlation
between the cut-off energy and spectral slope is inferred from the
{\it BeppoSAX} sample using {\sc pexrav} (see Matt 2001; Perola et al. 2002).

Independently, the narrow line component
detected by {\it Chandra} (Yaqoob et al. 2001), recently confirmed by
{\it XMM-Newton} (Pounds \& Reeves 2002), together with the constant 
line flux reported by Chiang et al. (2000) are suggestive of 
DCM dominated reflection in this source. 
Thus, although the close-by reflection model is not ruled out by
 the Compton bump data alone, the DCM picture appears favored when
these data are considered together with the iron line information.

\section{Discussion}

We considered the effects of reflection at large distance on the measured
reflection amplitude $R$ and equivalent width of the Fe fluorescence
line. We show that the remote cold material may in
principle account simultaneously for the observed correlation between $R$
and $\Gamma$, and the observed narrow line component which is predicted
to be independent of $\Gamma$.

We considered the $R$-$\Gamma$ relation produced for different fixed 
pivoting point energies, with and without varying 
the cut-off energy, in all cases 
a correlation is produced. 
Actually, in the case of a constant comptonised flux and
varying soft photon input, the pivot location is likely to fluctuate from
the fact that changes in the cooling also affects the cut-off energy, and
thus the energy range in which the comptonised power is released.  As a
consequence the pivot point may depend on $\Gamma$. Note that if the
spectral pivot shifts above 10-20 keV this produces an
anti-correlation. Such an anti-correlation is reported by Larmer
et al. (2000) in NGC 5506 where 
they find evidence for DCM reflection.
However, even if the pivoting point energy changes  
among different sources or within the 
spectral evolution of an individual source
we will just follow different correlations in the 
$R$-$\Gamma$ plane and generally, a correlation will be produced.
The correlation can be destroyed only if in a large fraction of the sources 
 the pivot energy decreases very quickly with $\Gamma$, 
or the pivot point is above a few 10 keV, in contrast to what
 is commonly observed. 

Thus, despite the uncertainties a correlation
 between $R$ and $\Gamma$ is 
expected in individual sources and is likely 
to be preserved in sample of sources.  
 In the latter case this interpretation requires 
 that most of the Seyfert sources present a similar variability mode
(i.e. pivoting primary power-law), time average properties. 
 as well as
geometry of the DCM.
While a pivoting powerlaw seems to be a general characteristics
 of Seyfert 1s, on the other hand, the geometry of the DCM is unknown.
According to the unified model, the geometry of the DCM 
should be similar in all sources. 
 In this context the small differences in geometry or
variability mode would explain part of the spread 
of the data points in the $R$-$\Gamma$ plane.

Our simulated $R$-$\Gamma$ correlation are qualitatively similar to
 the observed one, but not exactly identical. In particular it seems 
that many of the data points lie below the predictions in both Fig.~1 and 2.
This is not a fundamental problem for the DCM interpretation.
This is simply a confirmation that the simple pivoting powerlaw model 
is a poor approximation to the actual variability mode. 
For instance if the pivoting points depends on Gamma 
and if for some reasons 
the pivoting point shift toward energies below 2 keV 
for the hardest spectra, it can produce as many 
sources with low reflection as required.
However if we also consider the uncertainties on the measured $R$ 
 this would be certainly overinterpreting the data.
We thus prefered using a simple and approximative model
 that reproduces qualitatively 
the correlation, rather than a complicated and rather had hoc model.  

In this study we neglected the contribution of reflection from
 the central part of the accretion flow.
Actually the
 situation is probably more complex:
 the observed iron line profiles often present a broad
component which is unlikely to be produced far away from the central
engine. This broad component is probably associated with a inner 
reflection bump that may contribute
to the $R$-$\Gamma$ correlation.
It is however not clear if the broad iron line component $EW$
is correlated with $\Gamma$ as would be expected in such case 
(see LZ01 and 
Yaqoob et al. 2002 for a critical reevaluation of their results). 
At least, in several sources the broad line EW appears 
to be unrelated to the changes 
in $\Gamma$ and present complicated behavior 
(Nandra et al. 1999, Vaughan \& Edelson 2001 and references therein).

Another interesting point is the
the analogy with X-ray binaries where a similar  $R$-$\Gamma$ relations
is observed. Due to the time-scale involved this correlation
 cannot be accounted for
by a DCM reflector (Gilfanov, Revnitsev \& Churazov,  2000).
The shape of the observed correlation seems however to differ 
between X-ray binaries and Seyferts, in particular the DCM can 
naturally explain both the very large and very low reflection amplitude 
that are not observed in X-ray binaries. Also unlike AGNs, several X-ray
 binaries present a \emph{clear} iron line $EW$-$\Gamma$ correlation.
Actually accretion onto a black hole in an X-ray binary and an AGN is expected
to proceed in a very different way due to
 the different boundaries conditions at large distance.
 The correlation in both types of sources
 could thus have a different origin. 

In any case, the observations suggest
 that reflection arising from the DCM is important in many Seyfert galaxies.
In particular, we showed that in NGC 5548 the reflection component could be
dominated by the DCM reflector in the sense that, when uncertainties on
the cut-off energy are taken into account, the {\it RXTE} data of
 C00 are consistent with a constant reflected flux. This would solve the
previously reported inconsistency between the behaviour of the reflection
bump and the observed constant line flux. 

Finally we stress that our results do not exclude
 that part or even all of the $R$-$\Gamma$ correlation in Seyferts
 is due to some contribution from the central region, however they show 
that this is not required.
 This has important implications
 for the modeling of the inner part of the accretion flow. 
It demonstrates  
that in Seyfert galaxies, the $R$-$\Gamma$ correlation is not a good tool
to discriminate between models for the physics and geometry of 
the X-ray emitting region. In particular emission models
 that do not produce a $R$-$\Gamma$ correlation are not ruled out
 (see e.g. Malzac \& Celotti 2002).

\section{CONCLUSIONS}

In this paper we attempted to reconcile the numerous evidences
for DCM reflection in Seyfert 1s galaxies with the presence of
 a correlation between the reflection amplitude $R$ and the spectral
 index $\Gamma$. 
We found that in the case of reflection on a DCM,
 the generally observed variability mode
 in Seyfert 1s naturally leads to a $R$-$\Gamma$ correlation. 
We also showed 
that the measurement of $R$ and the shape of the
 observed $R$-$\Gamma$ correlation is sensitive to the value of 
the cut-off energy. As long as the cut-off energy cannot be constrained,
 this effect significantly increases the error bars on the measured $R$
 and possibly reduces the significativity of the correlation.
In our case study of NGC 5548, we showed that when the effects of the 
 cut-off energy are taken into account
 the data are fully 
consistent with the DCM picture,
 while in the context of close-by reflection 
the reported constant line flux remains a problem.

The EW of the narrow iron $K_{\alpha}$ line produced by reflection 
on the DCM is predicted to be generally independent of $\Gamma$.
This appears to be consistent with the present data, although the large 
uncertainties on the measurement of the the narrow line prevent us from 
reaching a firm conclusion.

In general, the possible presence of DCM reflection makes any attempt to
constrain the physical properties and the geometry of the central engine,
on the basis of the present reflection data, quite unreliable.
Further observational studies, with instruments having a broad band coverage
 ({\it INTEGRAL}) and high resolution around the iron line 
({\it Chandra}, {\it XMM-Newton} and latter {\it XEUS}),
 should allow one to disentangle the contribution 
of the DCM from that of the inner disc, and provide important 
insights on both the configuration of the DCM and the physics
 of the central engine.

\section*{Acknowledgments}
JM acknowledges grants from the Italian MURST (COFIN98-02-15-41) and the
European Commission (contract number ERBFMRX-CT98-0195, TMR network
"Accretion onto black holes, compact stars and protostars"). We thanks
James~Chiang who provided us with the NGC 5548 data as well as
Andrzej~Zdziarski and Cesare~Perola for the $R$-$\Gamma$ data.  We are
also indebted to Laura Maraschi for useful comments on the manuscript.

\end{document}